# How Often Should People be Tested for Corona to Avoid a Shutdown?

Helmut Hlavacs

University of Vienna, Faculty of Computer Science

helmut.hlavacs@univie.ac.at

April 27th 2020

## Abstract

Based on the well known SIR model, this paper develops a model for predicting the number of necessary testings of asymptomatic persons in order to push Reff below 1, thus suppressing an outbreak. The model considers R0, time for obtaining a test result, and effect of population discipline. The outcome are closed form expressions for the number of daily tests.

## Introduction

In light of the periodic nature of the Corona pandemic – even if eradicated, the Corona virus is likely to come back – the question arises how we can avoid a complete lockdown in the future, now that we know more about the spreading of this virus?

The major premise is the fact that we cannot build our actions based on herd immunity, since the percentage of people having contracted and survived the virus is much too small for this. Instead, we must strive to contain the next Covid-19 outbreak by ensuring that its effective reproduction number $R_{eff}$ is smaller than 1 from the get-go, to ensure that the epidemic dies out soon. This can be achieved by a highly disciplined population on the one hand, and rigorous mass testing of large parts of the population on the other hand, either through random sampling [14], group testing [15], or simply producing results faster. This paper aims at answering the question, how often people should be tested, even if they show no symptoms [12], in order to contain the epidemic without shutting down our economy? And once we have rough estimates, decide whether this is feasible at all, or not. The paper does not take into account sensitivity or specificity of COVID-19 tests [10], nor does it consider alternative lock-down ideas [11,17].

## Modelling the Spread

Basis for computing this number is the well know SIR model [1,2]. In this model, at any time $t \geq 0$, given a certain population of size $P$, $S(t) \leq P$ is the number of susceptible persons that potentially can contract the virus, $I(t)$ is the number of infected persons , and $R(t) \leq P$ is the number of recovered (or dead) persons. Hence, at any moment of time $t: S(t) + I(t) + R(t) = P$, with $S(0) = P - I_0, I(0) = I_0$, and $R(0) = 0$ for some small $I_0 > 0$. Basis for the determining the change over time of these variables is the *basic reproduction number* [9]

$$R_0 = c \, p \, D. \tag{1}$$

Here, $c$ denotes the average contacts a person has per day, $p$ is the probability of infecting one of these contacts, and $D$ is the period of communicability, i.e. the time an infected person can spread the virus. A person cannot spread the virus immediately after being exposed to it. Instead, communicability begins after some time called the latent period. From then on, the period of communicability begins and the person can spread the virus until immunity or death set in. According to [6] infectiousness starts roughly 2 days before symptom onset, and according to [7] ends 10 days after. Thus we estimate $D = 12$ on average.

After some time, more and more people may contract the virus, decreasing the number of susceptible people, and thus there are less people that can be infected by each infected person. This is represented by the effective reproduction number

$$R_{eff} = R_0 \frac{S(t)}{P} = c\,p\,D\,\frac{S(t)}{P}. \tag{2}$$

It is well known that once $R_{eff} < 1$, the infection eventually must die out.

## Measures against the Spread

We do not want the infection to grow rapidly, instead once it is detected it should be eradicated quickly, i.e. $S(t)$ should remain around $S(t) \approx S(0) \approx P$ and hardly change at all, and thus

$$R_{eff} \approx R_0 = c\,p\,D. \tag{3}$$

The first measure is *contact tracing*, which effectively reduces $R_{eff}$. [5] shows that this decrease grows nearly linearly with the percentage of traced contacts. E.g., if $R_0 = 3$ and 70% of the contacts are traced successfully, then $R_{eff}$ is roughly halve of $R_0$. The same is true for $R_0 = 4$ and 80% traced. In the present work this is factored in by assuming a lower $R_0 \in [1.5, 2.5]$, which usually is estimated in the range of $[1.4, 6.49]$, see [8].

From (3) it follows that we can further actively influence either $c$, $p$ or $D$. Influencing $c$ means reducing the number of physical contacts people have each day. This was the main measure against Covid-19 since the mitigation phase started. Effectively, the whole population was more or less quarantined. No doubt this had the desired effect on the spread, but with disastrous effects on our economy.

On the other hand, lowering $p$ to $p' = \alpha\,p$ for some $0 \leq \alpha < 1$ can be done by using for instance personal protective equipment like face masks, washing hands with soap regularly, or keeping physical distances between individuals in public spaces [13]. This was implemented during the mitigation phase, and without doubt had a large impact on the spread [16].

Finally, lowering $D$ to $D' = \beta\,D$ for some $0 \leq \beta < 1$ means quickly finding infectious persons and isolating them. The main tool for this is rigorous testing, here using PCR-based tests rather than anti-body tests. Since the number of such tests had to be ramped up starting at zero, testing so far was mainly concentrated on persons showing physiological symptoms like coughing or having difficulty to breathe. On the other hand, finding the large number of infectious persons showing no symptoms at all can only be done by testing large parts of the population on a regular basis.

Masks and physical distancing no doubt do influence the spread – if truly executed by the population. As this indeed can be observed at large scales in public places, supermarkets etc., we assume that we can reduce $p$ to $p' = \alpha\,p$ for some $0 \leq \alpha < 1$, the true value of $\alpha$ is of course unknown. In order to have some limits on $\alpha$ we assume that $\alpha\,R_0 > 1$, so that discipline alone is not enough for eradicating the virus, and therefore

$$\frac{1}{R_0} < \alpha < 1. \tag{4}$$

So finally, through testing and isolation, $D$ is reduced to $D' = \beta\,D$ for some $0 \leq \beta < 1$, resulting in

$$R'_{eff} \approx \alpha\,\beta\,c\,p\,D = \alpha\,\beta\,R_0. \tag{5}$$

In order to ensure $R'_{eff} < 1$, we must have $\alpha\,\beta < 1/R_0$ and because of (4) and (5)

$$\beta < \frac{1}{\alpha\,R_0} < 1. \tag{6}$$

## Modelling Infectiousness

We assume that people can get infected and turn infectious at some specific day. From then on they contribute to the infection for $1 \leq n \leq D$ further days. However, their degree of infectiousness may vary during this period, and we model this with a probability function

$$f(m) \geq 0, m \geq 1, \text{ with } f(m) = 0 \text{ for } m > D \text{ and } f(1) + \cdots + f(D) = 1. \tag{7}$$

If a person stays infectious for $n$ days then the *total contribution to spreading the disease* is

$$f(1) + \cdots + f(n) = F(n) \leq 1. \tag{8}$$

For computing $\beta$ we define a testing period of each person to have a length of $T > 0$ days. A person can turn infectious at any day $k$, $1 \leq k \leq T$, and will contribute to spreading the infection for $1 \leq n \leq D$ further days (at least one). Then we define the *reduction in spreading* the disease as

$$\beta(k) = F(n). \tag{9}$$

Especially, it takes $t \geq 0$ days to get the test result and isolate an infectious person. This means that if a *person turns infectious at the last day $T$ of the testing period*, then the person will have $t + 1$ days to spread the disease, and thus

$$\beta(T) = F(t+1). \tag{10}$$

At any day prior to this there is *one more day* to spread the disease, so

$$\beta(T-j) = F(t+1+j), 0 \leq j \leq T-1 \quad \text{or} \quad \beta(k) = F(T+t+1-k), 1 \leq k \leq T \tag{11}$$

and especially $\beta(1) = F(T+t)$. The *average reduction of spreading* the disease given $T$ and $t$ is then

$$\bar{\beta}(T) = \frac{1}{T}\sum_{k=1}^{T}\beta(k) = \frac{1}{T}\big(F(t+1) + \cdots + F(T+t)\big) \leq 1. \tag{12}$$

In the following we develop two versions of $F(n)$ and model their effect on the average $\bar{\beta}$ and determine the required testing period $T$ in order to ensure (6).

## Version 1: Uniform Infectiousness

In this scenario, we assume uniform infectiousness, i.e.

$$f(m) = \frac{1}{D}, 1 \leq m \leq D. \tag{13}$$

For summing up due to (8) we distinguish between two cases: (i) $T + t \leq D$, and (ii) $D \leq T + t$. Note that for $T + t = D$ the results are the same in both cases.

### Case $T + t \leq D$

In this case, if the person turns infectious at the first day $k = 1$ of the testing period then time of communicability is $T + t$ days and $\beta_1(1) = (T+t)/D$. If the person turns infectious at the last day $k = T$, the then time of communicability is $t + 1$ days and $\beta_1(T) = (t+1)/D$. In general

$$\beta_1(k) = \frac{T+t+1-k}{D}, \ 1 \leq k \leq T. \tag{14}$$

The average $\bar{\beta}_1(T)$ is then

$$\bar{\beta}_1(T) = \frac{1}{T}\sum_{k=1}^{T}\beta_1(k) = \frac{1}{TD}\big((t+1) + \cdots + (T+t)\big) = \frac{T+2t+1}{2D}. \tag{15}$$

## Case $D \leq T + t$

In this case, if the person turns infectious at day $k \leq T + t - D$ then testing does not shorten the period and $\beta_2(k) = 1$. If $k > T + t - D$ then the time of communicability is shortened and thus

$$\beta_2(k) = \begin{cases} 1 & \text{if} \quad k \leq T + t - D \\ \frac{T+t+1-k}{D} & \text{if} \quad T + t - D + 1 \leq k \leq T \end{cases} \quad (16)$$

The average $\bar{\beta}_2(T)$ is then

$$\bar{\beta}_2(T) = \frac{1}{T}\left(1 + \cdots + 1 + \frac{1}{D}((t+1)\ldots + D)\right) = \frac{T+t-D}{T} + \frac{(D-t)(D+t+1)}{2DT}. \quad (17)$$

It can easily be checked that in the border case $D = T + t$ it follows that $\bar{\beta}_1(T) = \bar{\beta}_2(T)$. Equ. (15) and (17) can now be combined to result in the average

$$\bar{\beta}(T) = \begin{cases} \bar{\beta}_1(T) & \text{if} \quad T + t \leq D \\ \bar{\beta}_2(T) & \text{if} \quad D < T + t \end{cases} \quad (18)$$

Figure 8 (left) shows the dependency of (12) on the parameters $D, t$ and $T$. Using this result we can now ask – given a certain goal $\bar{\beta}$, which testing period $T(\bar{\beta})$ is necessary to achieve it? In order to answer this question, we calculate the inverse functions of (15)

$$T_1(\bar{\beta}_1) = \bar{\beta}_1 \, 2 \, D - 2t - 1 \quad (19)$$

and (17)

$$T_2(\bar{\beta}_2) = \frac{(D-t)(D-t-1)}{2D(1-\bar{\beta}_2)}. \quad (20)$$

Following (18) we define

$$T(\bar{\beta}) = \begin{cases} \max\{T_1(\bar{\beta}), 0\} & \text{if} \quad T_1(\bar{\beta}) + t \leq D \\ \max\{T_2(\bar{\beta}), 0\} & \text{if} \quad D < T_1(\bar{\beta}) + t \end{cases} \quad (21)$$

The result is the inverse of (18) as shown in Figure 8 (right).

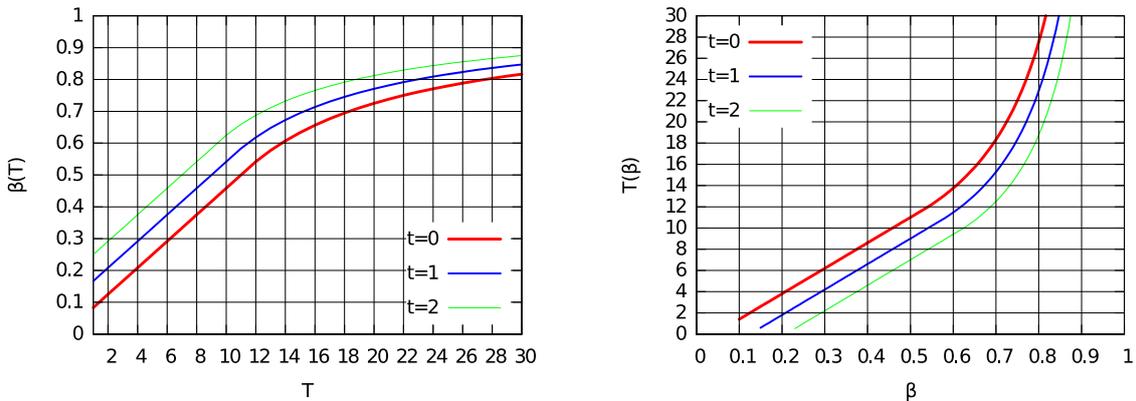

Figure 1: Average Beta depending on the testing period of T days (left), and its inverse function (right)

We can now answer the initial question. Given estimates for $\alpha$ and $R_0$ [5], we can use (6) and (21) to estimate the necessary maximal length of the testing period $T(\alpha)$. Note that the lower limit of $\alpha$ is limited by (6) and thus depends on $R_0$ (see Figure 2 and Figure 3). Low values of $\alpha$ represent a highly disciplined population, higher values an undisciplined population.

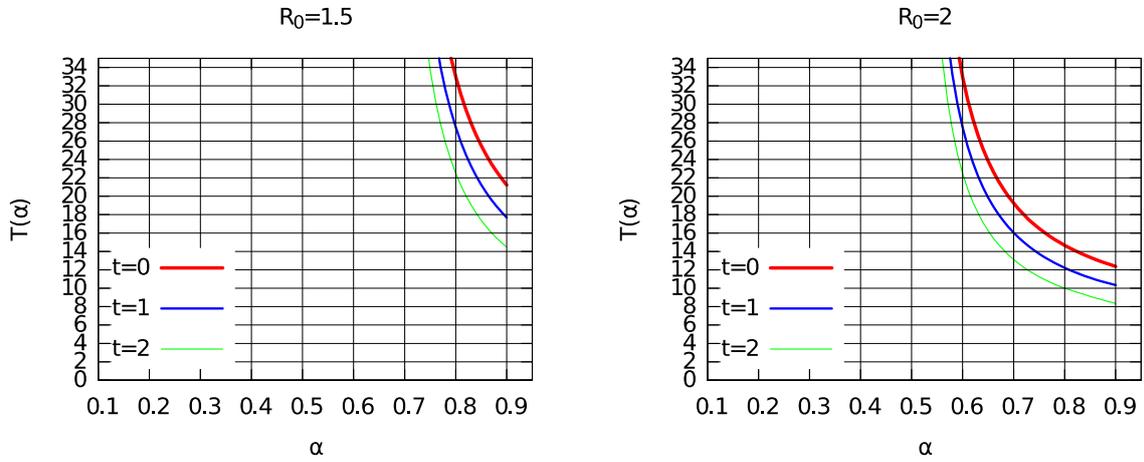

Figure 2: Max testing period of T days depending on population discipline for R0=1.5 (left), and R0=2 (right)

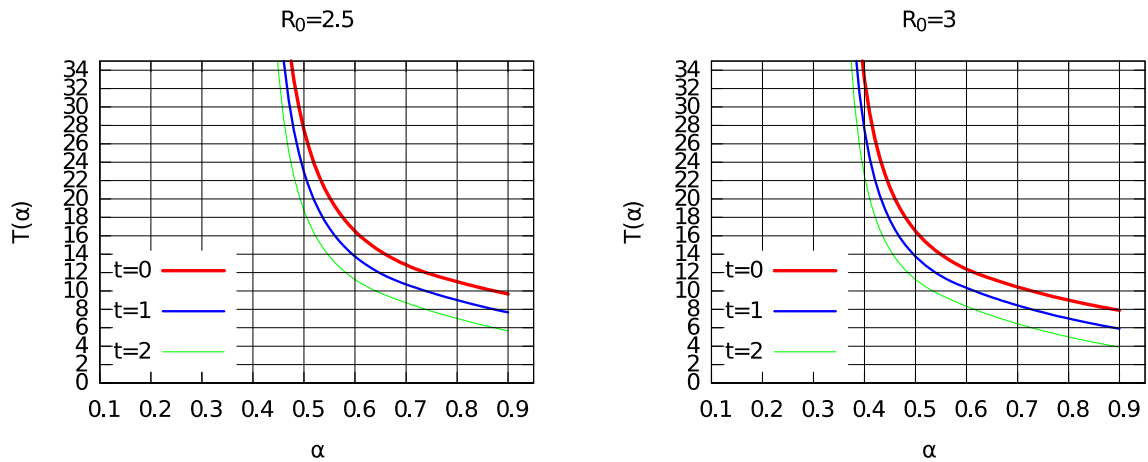

Figure 3: Max testing period of T days depending on population discipline for R0=2.5 (left), and R0=3 (right)

Since every person of a population should be tested every $T(\alpha)$ days, the number of *daily tests*

$$N(\alpha) = P/T(\alpha) \qquad (22)$$

for population size $P$. For instance, for Austria with $P = 8,000,000$, and disciplined population $\alpha = 0.8$ testing time $t = 1$ and efficient contact tracing resulting in $R_0 = 1.5$, every person should be tested about once every 28 days, resulting in around 286,000 tests per day. If results are available on the same day, then the period is increased to once per 33 days, or 242,000 tests per day.

Examples for this are shown in Figure 4, and Figure 5, with the same parameters as in Figure 2 and Figure 3. Figures show the necessary numbers of tests per day per Million inhabitants.

The curves shown in Figure 2, Figure 3, Figure 4, and Figure 5 reveal how important contact tracing and population discipline are. If both are effective then testing can be reduced drastically or even suspended. If people are undisciplined then the amount of necessary testing would be quite prohibitive.

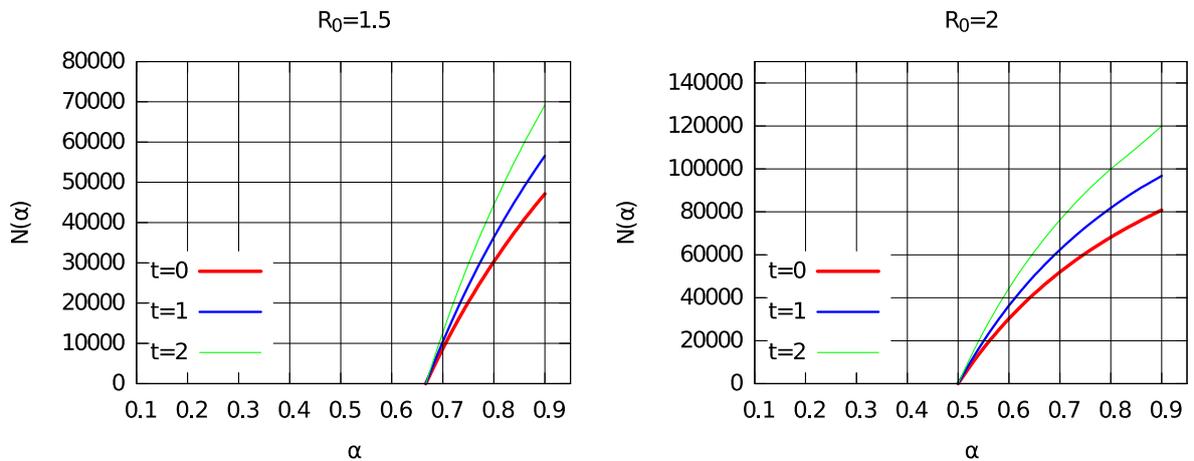

Figure 4: Tests per days per Million inhabitants depending on population discipline for R0=1.5 (left), and R0=2 (right)

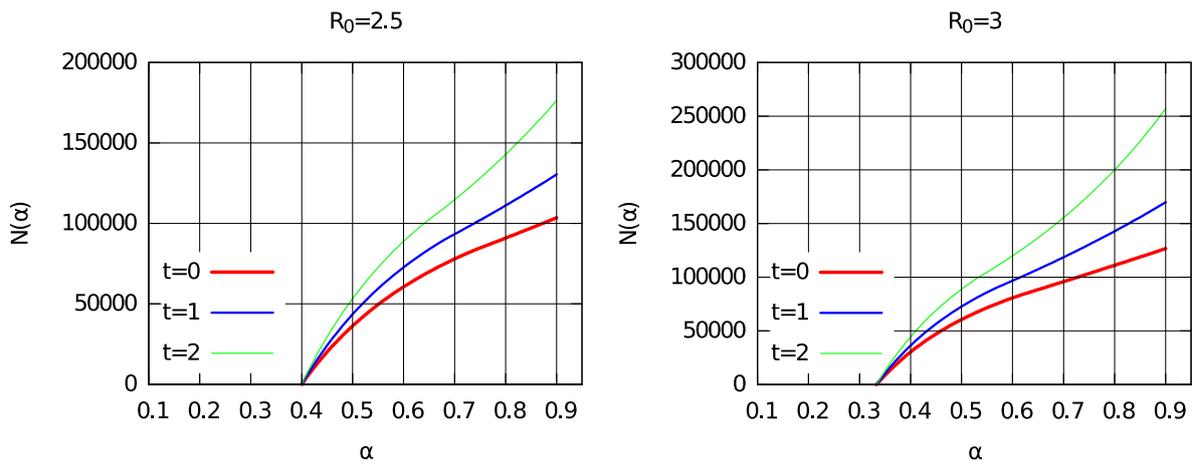

Figure 5: Tests per days per Million inhabitants depending on population discipline for R0=2.5 (left), and R0=3 (right)

## Version 2: Decreasing Infectiousness

In this version we assume that infectiousness is a result of time

For defining $\beta$ we assume that infections are passed on but that infectiousness is a function of time, as described in [6]. In [6] this dependency on time is modelled with gamma distributions, anchored around the time of symptoms onset. In the mentioned study, several cases where simulated and fit to measured data, conclusions were that the time of being infectious starts at least 2 days before symptom onset, and peaking between 2 and 1 days before. For this study we choose the inferred infectiousness profile as detailed in [6] Figure 1, starting 2 days before symptom onset and ending 10 days after [7]. We model this curve with a gamma distribution with scale 1.23 and rate 0.4, producing a density function as given in [6] Figure 1 (shifted by 2 days). Since this distribution is difficult to be handled analytically we approximate it with a Kumaraswamy distribution fulfilling the above mentioned criteria with parameters $a = 1$ and $b = 3$. A Kumaraswamy distribution is defined on the interval $[0,1]$ and has a density function of

$$f(x; a, b) = a\, b\, x^{a-1}(1 - x^a)^{b-1} \tag{23}$$

and a CDF of

$$F(x; a, b) = 1 - (1 - x^a)^b, \tag{24}$$

and especially

$$F(x; 1,3) = x^3 - 3x^2 + 3x. \tag{25}$$

We also scale its support from $[0,1]$ to $[0, D]$ and extend its range by defining

$$F(x) := \min\left\{F\left(\frac{x}{D}; 1,3\right), 1\right\} = \min\left\{\frac{x^3}{1728} - \frac{x^2}{48} + \frac{x}{4}, 1\right\}, x \geq 0. \tag{26}$$

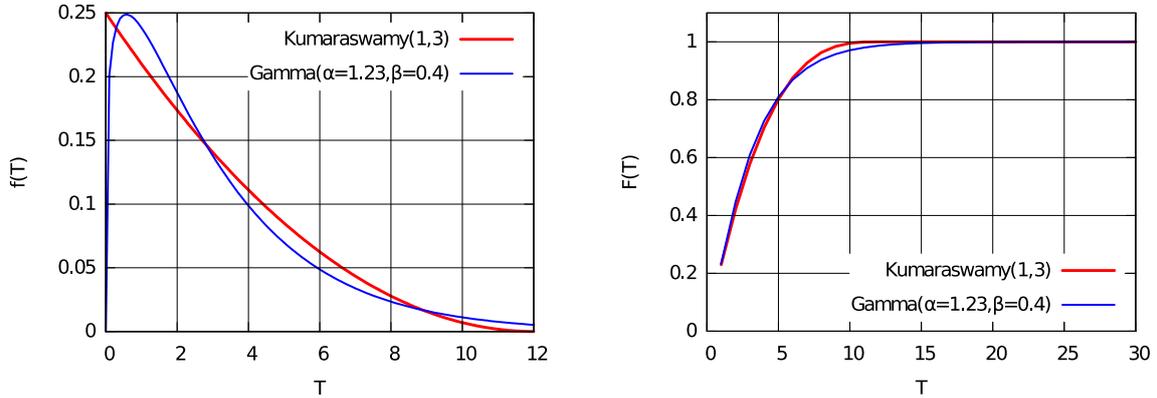

*Figure 6: Gamma and scaled Kumaraswamy density functions and CDFs, symptom onset at day 2*

Figure 6 shows that this is indeed a good fit, at least with respect to the cumulative distribution function (CDF). Both distributions have a mean of 3, and a probability mass before onset of ~0.4, the latter also reported in [6]. Since in our case the parameter $b = 3$ is an integer we can easily find an integral for the CDF, which is given as

$$G(x) = \int_0^x F(u)\, du = \begin{cases} G_1(x) = \frac{x^4}{6912} - \frac{x^3}{144} + \frac{x^2}{8} & \text{if } x \leq D \\ G_2(x) = G_1(D) + x - D = x - 3 & \text{else} \end{cases}. \tag{27}$$

Figure 7 shows the individual parts $G_1(x)$ and $G_2(x)$ and how they combine to $G(x)$. It also shows partial sums of (26), which are approximated by $G(x)$.

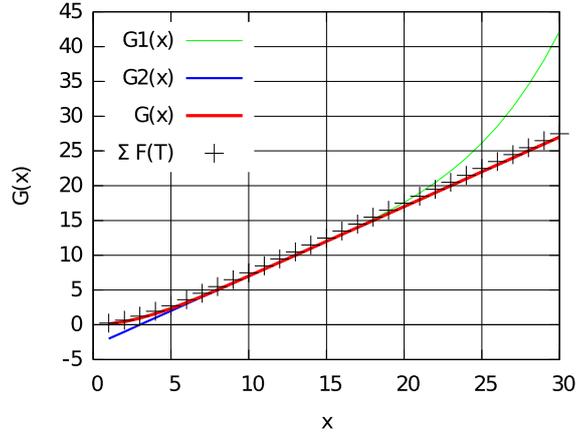

Figure 7: Integral function G(x) depending on G1(x) and G2(x), and summing F(x)

If an infectious person spreads the disease for $n > 0$ days, then the number of persons infected by this person is therefore on average smaller by a factor

$$\beta = F(n), n \geq 1. \tag{28}$$

The average $\bar{\beta}(T)$ is then

$$\bar{\beta}(T) = \frac{1}{T}\sum_{k=1}^{T}\beta(k) = \frac{1}{T}\big(F(t+1) + \cdots + F(T+t)\big). \tag{29}$$

Because there is no general explicit formula for this sum, we use (27) to estimate

$$\bar{\beta}(T) \approx \frac{1}{T}\int_{t+1-\frac{1}{2}}^{T+t+\frac{1}{2}} F(x)dx = \frac{1}{T}\Big[G\Big(T+t+\frac{1}{2}\Big) - G\Big(t+1-\frac{1}{2}\Big)\Big]. \tag{30}$$

Figure 8 (left) shows the dependency of $\bar{\beta}(T)$ on the parameters $t$ and $T$, using exact formula (29) and integral approximation (30).

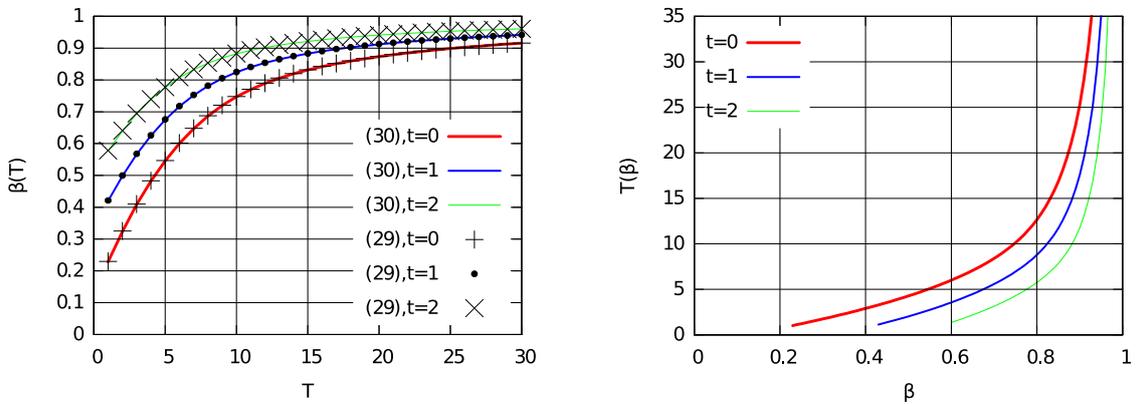

Figure 8: Average Beta depending on the testing period of T day (left), and inverse functions (right).

Since inverting (30) involves solving for the roots of a cubic polynomial, we can actually derive a closed form for its inverse. E.g., for $t = 0$ we derive

$$A_0 = \frac{\sqrt{5159780352\,\beta^2 - 8637594624\,\beta + 3625850161}}{4 \cdot 3^{3/2}}, \quad B_0 = \frac{(13824\,\beta - 1657) \cdot \frac{3}{2} - 36501}{6},$$

$$C_0 = \left(A_0 + B_0 + \frac{97336}{27}\right)^{1/3}, \quad T_{0,1}(\beta) = C_0 - \frac{529}{18\,C_0} + \frac{46}{3}, \quad T_{0,2}(\beta) = -\frac{279841}{110592\,\beta - 110592}.$$

$$T_0(\beta) = \begin{cases} \max\{T_{0,1}(\beta), 0\} & \text{if } T_{0,1}(\beta) + t \leq D \\ \max\{T_{0,2}(\beta), 0\} & \text{else} \end{cases} \tag{31}$$

For $t = 1$ we derive

$$A_1 = \frac{3^{3/2}\sqrt{7077888\,\beta^2 - 12399616\,\beta + 5439377}}{4}, \quad B_1 = \frac{(13824\,\beta - 4563) \cdot \frac{3}{2} - 27783}{6},$$

$$C_1 = (A_1 + B_1 + 2744)^{1/3}, \quad T_{1,1}(\beta) = C_1 - \frac{49}{2\,C_1} + 14, \quad T_{1,2}(\beta) = -\frac{7203}{4096\,\beta - 4096}$$

$$T_1(\beta) = \begin{cases} \max\{T_{1,1}(\beta), 0\} & \text{if } T_{1,1}(\beta) + t \leq D \\ \max\{T_{1,2}(\beta), 0\} & \text{else} \end{cases} \tag{32}$$

and for $t = 2$ we derive

$$A_2 = \frac{\sqrt{5159780352\,\beta^2 - 9371372544\,\beta + 4258638073}}{4 \cdot 3^{3/2}}, \quad B_2 = \frac{(13824\,\beta - 6965) \cdot \frac{3}{2} - 20577}{6},$$

$$C_2 = \left(A_2 + B_2 + \frac{54872}{27}\right)^{1/3}, \quad T_{2,1}(\beta) = C_2 - \frac{361}{18\,C_2} + \frac{38}{3}, \quad T_{2,2}(\beta) = -\frac{130321}{110592\,\beta - 110592}.$$

$$T_2(\beta) = \begin{cases} \max\{T_{2,1}(\beta), 0\} & \text{if } T_{2,1}(\beta) + t \leq D \\ \max\{T_{2,2}(\beta), 0\} & \text{else} \end{cases} \tag{33}$$

Results for these inverted functions are depicted in Figure 8 (right).

Figure 9 and Figure 10 show the connection between population discipline $\alpha$, testing delay $t$, and testing period $T$. Compared to the uniform version, results require much more tests here. For instance, in the Austrian case ($P = 8,000,000$), for $R_0 = 1.5$, $\alpha = 0.8$ and $t = 1$ the testing period can at most be 10 days per person, resulting an 800,000 tests per day. In case tests are done faster, i.e. $t = 0$, the testing period can at most be 15 days, or 533,000 tests per day.

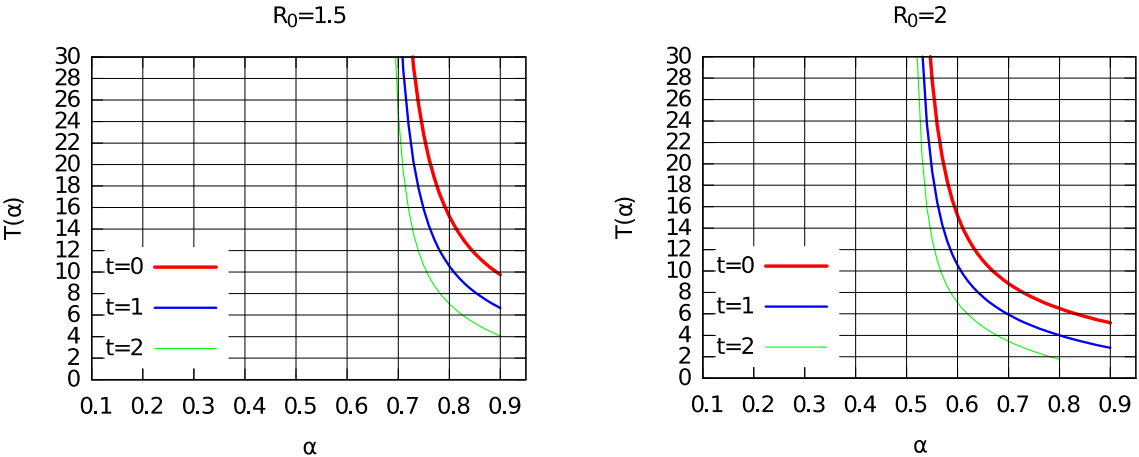

Figure 9: Max testing period of T days depending on population discipline for R0=1.5 (left), and R0=2 (right)

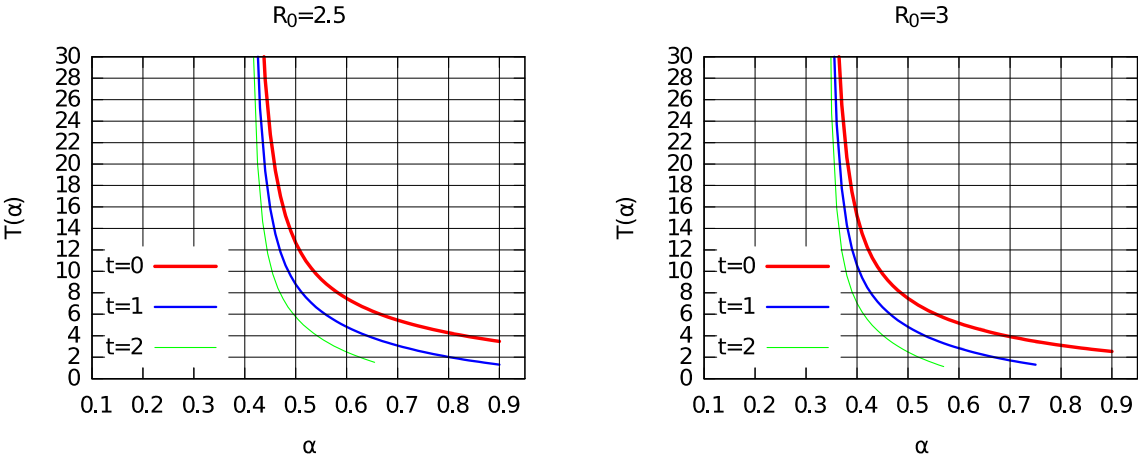

Figure 10: Max testing period of T days depending on population discipline for R0=2.5 (left), and R0=3 (right)

Likewise, Figure 11 and Figure 12 show the number of necessary tests per Million inhabitants.

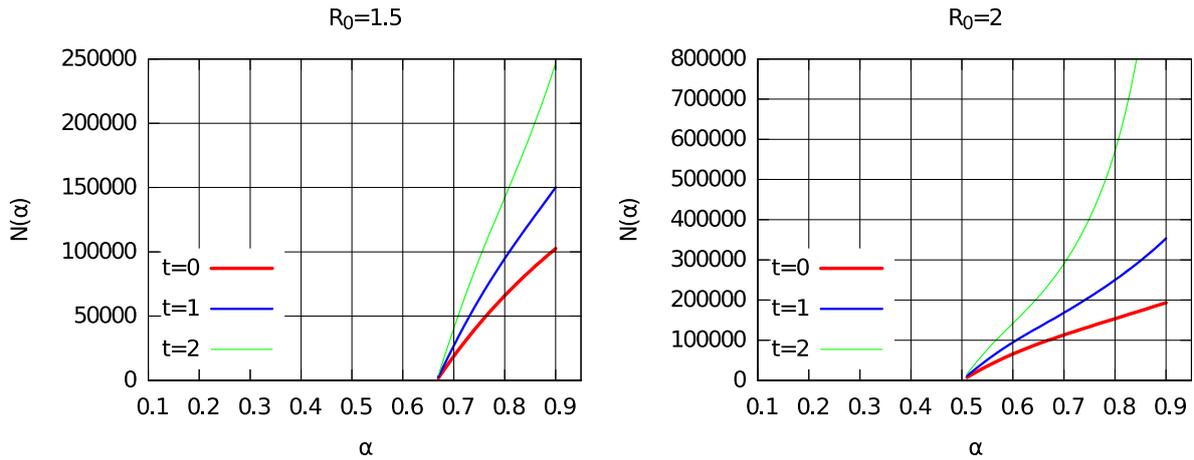

*Figure 11: Tests per days per Million inhabitants depending on population discipline for R0=1.5 (left), and R0=2 (right)*

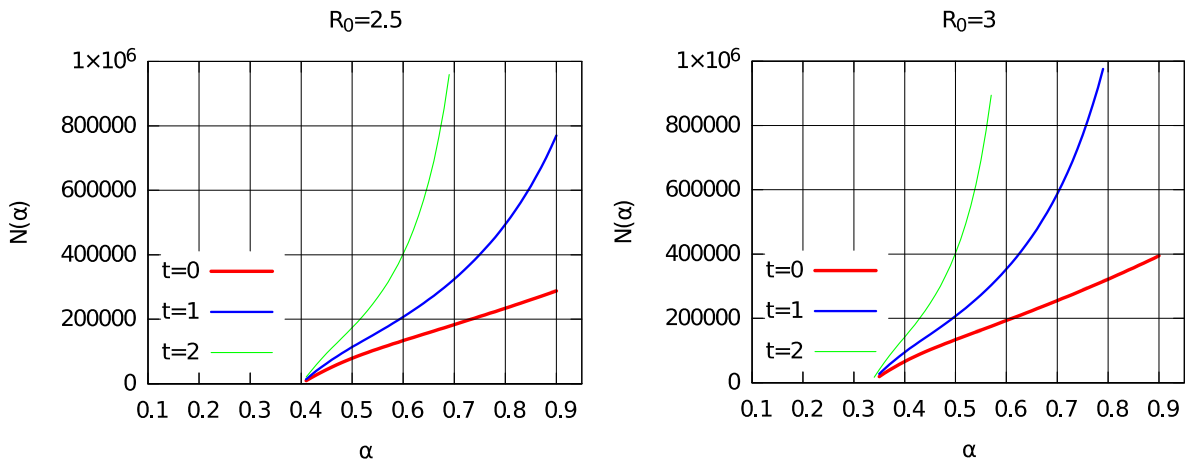

*Figure 12: Tests per days per Million inhabitants depending on population discipline for R0=2.5 (left), and R0=3 (right)*

Again the influences of the main parameters $\alpha$, $t$ and $R_0$ are clearly visible. If contact tracing is highly efficient, a little more discipline can actually push decrease the number of tests significantly.

### Discussion of the Results

The results prove to cover a wide range of possible outcomes. In particular, we see the influence of $\alpha$ on the scales. There is a region of instability for $\alpha$, and if $\alpha$ comes near this region then results quickly get much better, increasing the required testing period significantly. However, the farther away $\alpha$ gets from this regions, the number of required daily tests quickly rises.

The starting point is thus somewhat instable, and any deviation of the population from "ideal" behavior is quickly translated in a sharp increase of necessary tests. After this, the time between taking the test and isolation $t$ has a severe impact on the results. Basically the quicker the better.

The basic reproduction number $R_0$ shifts the curves horizontally to the left. The results indicate that decreasing $R_0$ has a very positive effect on the number of needed tests.

On the other hand, in pessimistic scenarios, with high reproduction numbers, unsuccessful tracing, long testing waiting times and an undisciplined population, the amount of tests necessary is clearly unrealistic, at least for the time being.

However, results show that if we can further motivate the population to keep physical distance outside, wear face masks permanently, avoid physical contact with people from other households, wash their hands with soap regularly, and do not touch their faces outside (all of which influence the value of $\alpha$), and furthermore there is a rigorous implementation of testing and contact tracing, then the spread can indeed be contained without further shutdowns, just by implementing rigorous testing on a massive scale.

## Conclusion

This paper discusses how many daily tests are necessary in order to contain Covid-19 without shutting down our economy. Modeling outcomes show that this is indeed possible, at the high cost of a large number of tests to be carried out on a daily basis, rigorous contact tracing and a highly disciplined population. If only one of these factors is not given, then an economic shutdown similar to the first one in March 2020 is probably the only choice left.

The presented results are not meant to yield exact figures. Rather they should indicate the order of magnitudes as well as the dependencies between the parameters and the required testing. Results can now act as guidelines to estimate the amount of testing necessary to achieve the goal of containment.

## Acknowledgements

I would like to thank Niki Popper and colleagues for their valuable comments.